\documentclass[showkeys,showpacs,prb,twocolumn]{revtex4-1}
\usepackage{amsmath}
\usepackage{color}
\usepackage{graphicx}
\usepackage{setspace}
\usepackage{comment} 

\newcommand{\GP}{{Gr\"uneisen parameter}}

\newcommand{\mos}{{MoS$_2$}}
\newcommand{\Gru}{{Gr\"uneisen}}
\newcommand{\VEC}[1]{{\boldsymbol{ #1}}}
\newcommand{\invcm}{cm$^{-1}$}

\begin{document}
\title{
Direct calculation of the linear thermal expansion coefficients of \mos{} via symmetry-preserving deformations
}

\author{Chee Kwan Gan}
\email{Corresponding author: ganck@ihpc.a-star.edu.sg}
\affiliation{Institute of High Performance Computing, 1 Fusionopolis Way, \#16-16 Connexis, Singapore 138632}
\author{Yu Yang Fredrik Liu}
\affiliation{Institute of High Performance Computing, 1 Fusionopolis Way, \#16-16 Connexis, Singapore 138632}

\date{7 October 2016}

\begin{abstract}
Using density-functional perturbation theory and the Gr\"uneisen formalism, we directly calculate the linear thermal expansion
coefficients (TECs) of a hexagonal bulk system \mos{} in the crystallographic $a$ and $c$ directions.
The TEC calculation depends critically on the evaluation of a temperature-dependent quantity $I_i(T)$,
which is the integral of the product of heat capacity and $\Gamma_i(\nu)$, of frequency $\nu$ and strain type $i$, where
$\Gamma_i(\nu)$ is
the phonon density of states weighted by the \GP{}s.
We show that to determine the linear TECs we may use 
minimally two uniaxial strains in the
$z$ direction, and either the $x$ or $y$ direction. 
However, 
a uniaxial strain in either the $x$ or $y$ direction
drastically reduces the symmetry of the
crystal from  a hexagonal one to a base-centered orthorhombic one.
We propose to use an efficient and accurate symmetry-preserving
biaxial strain in the $xy$ plane
to derive the same result for $\Gamma(\nu)$.
We highlight that the \GP{} associated with a biaxial strain may not be the same 
as the average of \GP{}s associated with two separate 
uniaxial strains in the $x$ and $y$ directions 
due to possible preservation of degeneracies of the phonon modes under a biaxial deformation.
Large anisotropy of TECs is observed where the linear TEC in the $c$ direction
is about $1.8$ times larger than that in the $a$ or $b$ direction at high temperatures.
Our theoretical TEC results are compared with experiment. The symmetry-preserving approach
adopted here may be applied to a broad class of two lattice-parameter systems such
as hexagonal, trigonal, and tetragonal systems, which
allows many complicated systems to be treated on a first-principles level.
\end{abstract}


\maketitle

\section{Introduction}

Transition-metal dichalcogenides (TMDs) $TX_2$, where $T$ is a transition metal (such as W and Mo) and $X$ is
a chalcogen (such as S, Se, and Te), receive considerable attention due to their important mechanical and electronic
 properties\cite{Wilson69v18}. 
Molybdenum disulfide (\mos), a prototypical example of TMDs, is a layered system
where Mo atoms form hexagonal layers\cite{Verble70v25,Wieting71v3}. Each of the Mo hexagonal layers
is sandwiched between two similar lattices of S atoms, forming a trilayer\cite{Zhang13v87,Zhao13v13}.
The atoms within each trilayer are held together by strong covalent bonds, 
while the trilayers of \mos{} interact primarily through
weak van der Waals interactions. 
It is this sandwiched structure that endows \mos{} with the
important mechanical properties
for solid lubricants\cite{Aksoy06v67,Lee10v328}. The electronic, 
optical, and lattice dynamical properties have been 
under intense investigations\cite{Verble70v25,Wieting71v3,Bhatt14v45}.
The research on multilayers of \mos{}, among many other multilayers of TMDs,
has been fueled by their novel properties intrinsic to two-dimensional materials. For example, successes of \mos{} multilayers
have been demonstrated for the purposes of energy-efficient field-effect transistor\cite{Radisavljevic11v6}, 
advanced electrocatalysts\cite{Li11v133}, thermoelectric devices\cite{Buscema13v13,Huang14v16} with a large and tunable Seebeck coefficient,
phototransistors\cite{Yin11v6}, superconductivity\cite{Costanzo16v11}, etc. 
\mos{} is joining the ranks
of other low-dimensional materials, demanding both efficient and accurate treatment
of a first-principles approach\cite{Molina-Sanchez11v84,Cheng12v2,Cai14v89,Sevik14v89}.
Even though the mechanical, electronic, and lattice dynamical 
properties of the equilibrium structure of
\mos{} have been studied extensively\cite{Aksoy06v67,Lee10v328,Ataca11v115}, there are
relatively few first-principles studies of the anharmonic effects\cite{Shiomi11v84} that 
contribute to the thermal properties such as thermal 
conductivity and thermal expansion coefficient (TEC).

The linear TECs of 2H-\mos{} have been measured in
Refs.~\onlinecite{El-Mahalawy76v9,Murray79v12} where it was found that 
the TEC along the $c$ direction
is larger than that along the $a$ direction.
On the theoretical side, TECs may be calculated by
solving the vibrational self-consistent field equations\cite{Monserrat13v87} or the 
nonequilibrium Green's function method\cite{Jiang09v80}.
TECs may also be determined from a quasiharmonic 
approximation (QHA) calculation
in which a set of calculations is to be carried out over
a grid or mesh of lattice-parameter points, where the dimensionality of the 
grid
depends on the number of independent lattice parameters\cite{Lazzeri98v81,Mounet05v71}.
Recently Ding and Xiao\cite{Ding15v5} chose six volumes to perform 
phonon calculations to first obtain the volumetric TEC. Another relation involving 
the linear TECs for $a$ and $c$ [Eq.~15 of Ref.~[\onlinecite{Ding15v5}]] was set up, and the values of TECs were solved.
In this work, we develop a direct approach based on the Gr\"uneisen formalism
to calculate the TECs in the $a$ and $c$ directions.
Our TEC results are then compared with experiment.
The outline of this paper is as follows: Section~\ref{sec:method} discusses the 
methodology used to efficiently calculate the thermal expansion coefficients of a general
hexagonal system. Section~\ref{sec:results} reports the results and discussion of the application of the
method to \mos{}. Section~\ref{sec:conclusions} contains the conclusions.

\section{Methodology}
\label{sec:method}

We shall first present the expressions for 
TECs for a general hexagonal system obtained with the
\Gru{}
formalism.\cite{Gruneisen26v10,Pavone93v48,Barron80v29,Schelling03v68,Gan15v92}
Results specific to the hexagonal \mos{} will be presented later.
The linear TECs of the crystal along the $x$, $y$ and $z$ directions,
denoted by $\alpha_{1}$, $\alpha_{2}$, and $\alpha_{3}$,
at temperature $T$ can be described by a matrix equation
\begin{equation} 
\begin{pmatrix} \alpha_{1}\\ \alpha_{2}\\ \alpha_{3}
\end{pmatrix} = \dfrac{1}{\Omega}
C^{-1} \begin{pmatrix} I_{1}\\ I_{2}\\
I_{3} \end{pmatrix}
\label{eq:alpha}
\end{equation} where $\Omega$ is the equilibrium
volume of the primitive cell and $C^{-1}$ is the elastic compliance
matrix\cite{Kittel96}. The values $C_{ij}$ are the matrix elements of the
elastic constant matrix $C$ that 
corresponds to a hexagonal system\cite{Nye-book} where
\begin{equation}
C = \begin{pmatrix}
C_{11} & C_{12} & C_{13} \\
C_{12} & C_{11} & C_{13} \\
C_{13} & C_{13} & C_{33} \\
\end{pmatrix}
\end{equation}
The integrated quantities in Eq.~\ref{eq:alpha} are given by
\begin{equation}
I_{i}(T) = \dfrac{\Omega}{(2\pi)^{3}}
\sum_{\lambda} \int_{\rm BZ} d\VEC{k}\ \gamma_{ i, \lambda\VEC{k} } c( \nu_{\lambda
\VEC{k}},T)
\label{eq:Integ}
\end{equation}
where the integral is over the first Brillouin
zone (BZ).  The frequency $\nu_{\lambda\VEC{k}}$ of a phonon mode depends on the mode
index $\lambda$ and wavevector $\VEC{k}$.  The heat capacity contributed by a
phonon mode with frequency $\nu$ at temperature $T$ is $ c(\nu,T) = k_{B}
(r / \sinh r)^2$ with $ r = h \nu / 2 k_{B} T $, where $h$ and $k_{B}$
are the Planck and Boltzmann constants, respectively.  The \GP{}
$ \gamma_{ i, \lambda\VEC{k} }  = - \nu_{\lambda \VEC{k}}^{-1}
\partial  \nu_{\lambda \VEC{k}} / \partial \epsilon $ measures
the relative change of a phonon frequency $\nu_{\lambda \VEC{k}}$ as
a result of an $i$-type deformation with strain size $\epsilon$ applied to 
the crystal.
For example, if a uniaxial strain is applied in the $x$ direction,
then the strain parameters are $(\epsilon,0,0,0,0,0)$ (in the Voigt 
notation\cite{Nye-book}), i.e., $\epsilon_1 = \epsilon$, and $\epsilon_j = 0$, for $j = 2, \cdots, 6$.
We apply uniaxial strains in the $x$, $y$, and $z$ directions to determine $I_1(T)$, $I_2(T)$, and $I_3(T)$, respectively.
\GP{}s are evaluated using a central-difference scheme, where
a change in the dynamical matrices before and after
deformation is used in the perturbation theory to 
deduce the changes in eigenfrequencies\cite{Gan15v92}.

By a proper sampling in the $k$-space, we may calculate the phonon density of states
as 
\begin{equation}
\rho(\nu) = \frac{\Omega}{(2\pi)^{3}} \sum_{\lambda} \int_{\rm BZ} d\VEC{k}\ \delta(\nu - \nu_{\lambda\VEC{k} })
\end{equation}

Next we introduce a related quantity, $\Gamma_{i}(\nu)$, the
phonon density of states weighted by the \GP{}s as
\begin{equation}
\Gamma_{i}(\nu) =
\dfrac{\Omega}{(2\pi)^{3}} \sum_{\lambda} \int_{\rm BZ} d\VEC{k}\ \delta(\nu
- \nu_{\lambda\VEC{k} }) \gamma_{ i, \lambda\VEC{k} } 
\label{eq:Gnu}
\end{equation}
The usefulness of $\Gamma_i(\nu)$ is that we may obtain $I_i(T)$ in Eq.~\ref{eq:Integ} from another
relation
\begin{equation}
I_i(T) = \int_{\nu_{\rm min}}^{\nu_{\rm max}} d\nu\ \Gamma_i(\nu) c(\nu,T)  
\label{eq:Integ2}
\end{equation}
where $\nu_{min}$ and $\nu_{max}$ are the minimum and maximum frequencies of
all phonon modes in the BZ, respectively.

To calculate the linear TECs, it appears that a
set of three uniaxial deformations in the $x$, $y$, and $z$ directions 
is needed. However, 
due to the symmetry of the hexagonal system, we should have  $\alpha_1 = \alpha_2$ on 
physical grounds and hence
$I_1(T) = I_2(T)$, so that
the TEC Eq.~\ref{eq:alpha} reduces to 
\begin{equation} 
\begin{pmatrix} \alpha_{1}\\ \alpha_{3}
\end{pmatrix} = \dfrac{1}{\Omega   } \begin{pmatrix}
[ C_{11} + C_{12} ]& C_{13} \\
2C_{13} & C_{33} \\
\end{pmatrix}^{-1} \begin{pmatrix} I_{1}\\
I_{3} \end{pmatrix}
\label{eq:2d1}
\end{equation} 
or
\begin{equation} 
\begin{pmatrix} \alpha_{1}\\ \alpha_{3}
\end{pmatrix} = \dfrac{1}{\Omega   D} \begin{pmatrix}
C_{33} & -C_{13} \\
-2C_{13} & [C_{11} +C_{12}] \\
\end{pmatrix} \begin{pmatrix} I_{1}\\
I_{3} \end{pmatrix}
\end{equation} 
where $D = (C_{11} + C_{12}) C_{33} - 2C_{13}^2$.
Therefore, for a hexagonal system  two uniaxial strains, the first one in either 
the $x$ or
$y$ direction and the second one in the $z$ direction, are sufficient to determine the
linear TECs. For \mos{}, the symmetry of the hexagonal system is not altered [the space group remains as
$P6_3/mmc$ $(\# 194)$] when a 
uniaxial strain is applied in the $z$ direction.
However, the symmetry is significantly lowered from hexagonal with a space group of $P6_3/mmc$ $(\# 194)$
 to base-centered orthorhombic with a space group of $Cmcm$ $(\# 63)$ after a uniaxial strain
is applied in the $x$ or $y$ direction.
This will result in an increase of the computational cost compared
to that which preserves the hexagonal symmetry where a phonon calculation
is to be performed.  
For example, after
applying a uniaxial strain in the $x$ direction, a $5\times 5 \times 5$ $q$ mesh required in a phonon calculation will result in
$27$ irreducible $q$ points and $486$ irreducible representations or $486$
self-consistent field calculations. 
This is to be compared with the
symmetry-preserving deformations (e.g., a uniaxial strain in the $z$ direction)
where the number of irreducible $q$ points is $15$ and 
the number of irreducible representations is $252$, which 
clearly shows substantial computational savings. More savings are
expected when 
complicated crystal structures are treated.

We propose to use a computationally efficient, 
symmetry-preserving biaxial strain in the $xy$ plane 
(hereafter it shall be called an $xy$ biaxial strain)
where $\epsilon_1 = \epsilon_2 = \epsilon $
to evaluate the \GP{}s $ \gamma_{b, \lambda\VEC{k} } $
and use Eq.~\ref{eq:Integ} or Eq.~\ref{eq:Integ2} to obtain $I_{b}(T)$. 
Due to the underlying symmetry, 
 $I_{b}(T) = I_1(T) = I_2(T)$.
However it should be noted that \GP{}s
due to an $xy$ biaxial strain may not be the same as the average
of the \GP{}s due to $x$ and $y$ uniaxial strains. These points will
be elaborated later. 

The phonon spectra of \mos{} are calculated with density functional
perturbation theory (DFPT)\cite{Baroni01v73}.
For the unstrained structure, a $q$ mesh of $5 \times 5 \times 5$ is used for the phonon calculations,
which is equivalent to evaluating the force constants\cite{Liu14v16,Gan06v73}
using a $5 \times 5 \times 5$ supercell.  
The phonon calculations proceed by evaluating dynamical matrices at a 
number of irreducible $q$ points.
From the dynamical matrices the interatomic force
constants in the real space 
are obtained by an inverse Fourier transform, and these force constants are
used to construct dynamical matrices at any $\VEC{k}$ to calculate 
the phonon eigenfrequencies
$\nu_{\lambda\VEC{k}}$.
For the strained structures, a $q$ mesh of $5\times 5 \times 5$ is also used. For the unstrained structure,
a larger $q$ mesh of $6\times 6 \times 6$ is used to confirm that a $q$ mesh of $5 \times 5 \times 5$ is sufficient for the 
purposes of TEC calculations.

\section{Results and Discussions}
\label{sec:results}

The bulk \mos{} belongs to the $P6_{3}/mmc$ nonsymmorphic space group (\#~$194$), 
with two inequivalent atoms, where a Mo atom occupies a 
$ 2c(1/3,2/3, 1/4)$ site and 
a S atom occupies a $4f(1/3,2/3,z)$ site, $z = 0.6213$. This
gives a total of six atoms in the hexagonal primitive cell.  Density-functional
theory (DFT) calculations are carried out using the plane-wave basis code
Quantum Espresso\cite{Giannozzi09v21}. The orientation of the crystal
adopted in this work is dictated by the choice of the primitive lattice 
vectors, where $\VEC{a}_1 = (a,0,0)$, $\VEC{a}_2 =
(-a/2,a\sqrt{3}/2,0)$, and $\VEC{a}_3 = (0,0,c)$ and $a$ and $c$
are the hexagonal lattice parameters.
We use $60$~Ry as the cutoff
energy for the plane-wave basis set. The local-density
approximation (LDA) is used to describe the exchange and correlation.
For \mos{} multilayers it has been demonstrated\cite{Zhao13v13} that the calculated
phonon frequencies using LDA agree well with the experimental
results. However, we should point out that the van der Waals interactions
may be important in the calculation of phonon dispersion relations for
some layered systems such as graphite\cite{Sabatini16v93}.
Pseudopotentials for Mo and S are generated from pslibrary.1.0.0
based on the Rappe-Rabe-Kaxiras-Joannopoulos\cite{Rappe90v41}
scheme. A $13 \times 13 \times 4$ Monkhorst-Pack
$k$-point mesh is used. The hexagonal
lattice parameters and the atomic positions are fully relaxed.
The force tolerance is taken to be
$10^{-3}$~eV/\AA.  We obtain $a_{0} = 3.125$~\AA{} and $c_{0} = 12.086$~\AA,
which is in good agreement with the experimental result\cite{Boker01v64} of $a = 3.160$~\AA, and $c= 12.294$~\AA.
This is also consistent with the fact that LDA tends to overbind in crystals.
We perform elastic-constant
calculations\cite{Beckstein01v63,DalCorso16v28} to obtain the elastic constants $(C_{11},
C_{12}, C_{13}, C_{33}, C_{44}) = (242.35, 58.84, 11.31, 51.70, 19.60)$~GPa.
These results are in very good agreement with
other computational results\cite{Peelaers14v118} where the values of $(C_{11},
C_{12}, C_{13}, C_{33}, C_{44}) = (238, 64,12,57,18)$~GPa. The
agreement of our results with the experimental results\cite{Feldman76v37} is
rather good except for $C_{12}$ and $C_{13}$ (where the experimental values
of $C_{12} = -54$~GPa and $C_{13}=23$~GPa), which may be due to
the fact these two values are not directly determined in
experiment, as discussed in Ref.~[\onlinecite{Peelaers14v118}].

\begin{figure}
\includegraphics[clip,width=8.0cm]{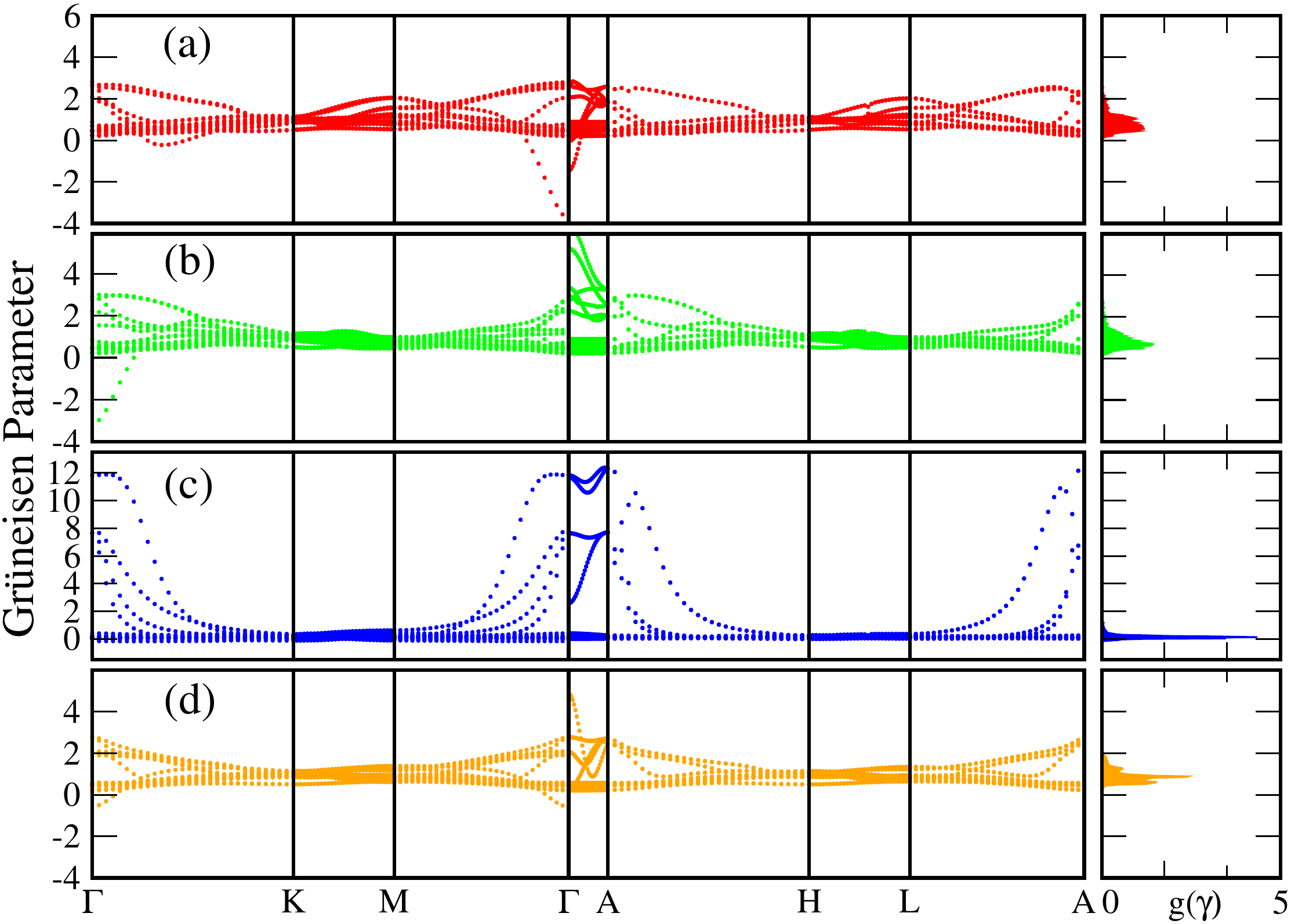}
\caption{
(Color online) 
\GP{}s $\gamma_{i, \lambda\VEC{k}}$ along the high-symmetry directions for hexagonal \mos{}, for 
(a) $x$ uniaxial, (b) $y$ uniaxial, (c) $z$ uniaxial, and (d) $xy$ biaxial strains.
The corresponding densities of \GP{}s, $g_i(\gamma)$, are shown on the right.
A mesh of $30\times 30 \times 10$ for the $k$-point sampling
is used to calculate $g_i(\gamma)$.
}
\label{fig:GP}
\end{figure}

\begin{figure}
\includegraphics[clip,width=8.0cm]{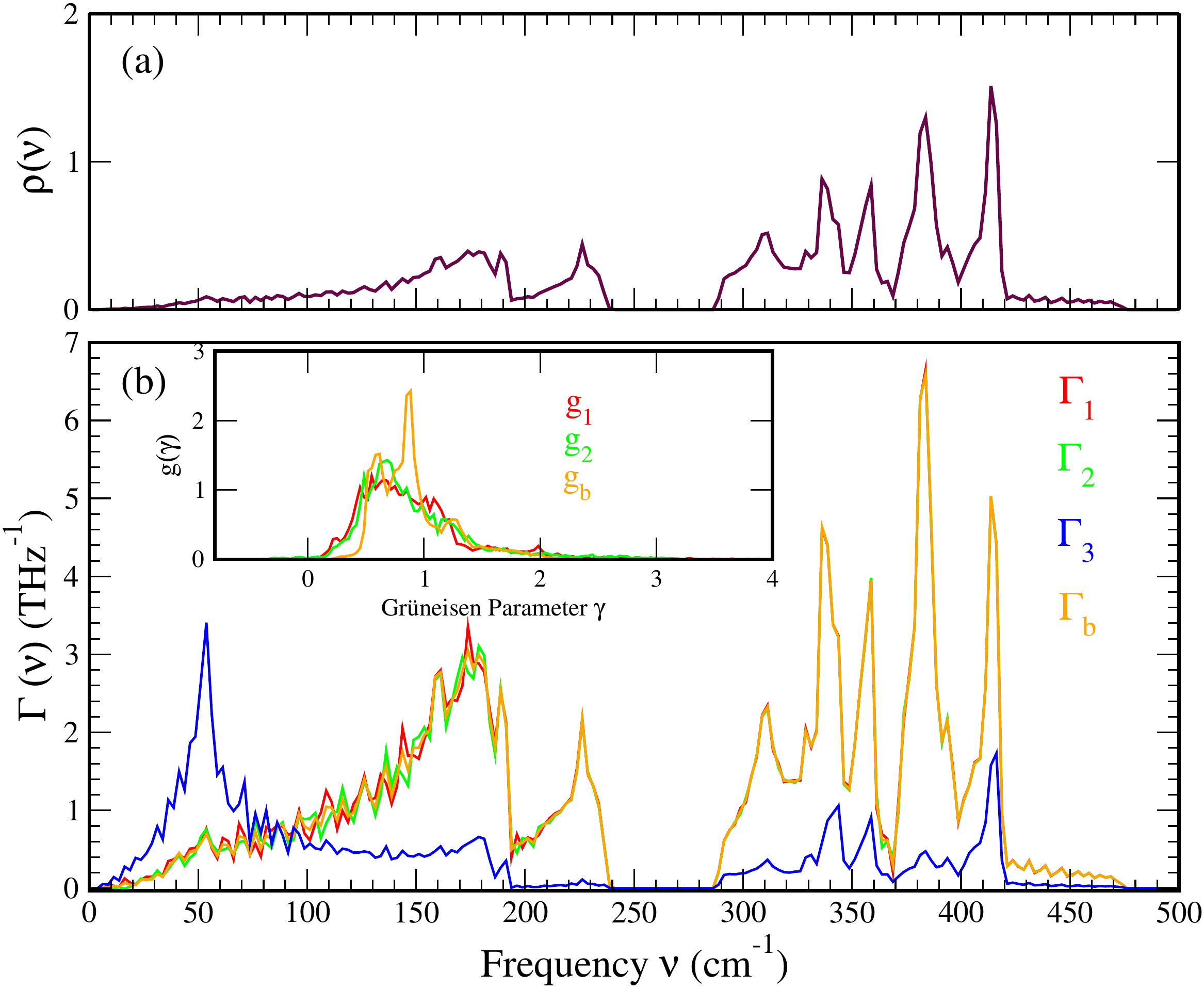}
\caption{
(Color online)
(a) The phonon density of states $\rho(\nu) $ for the unstrained \mos{} 
structure.
(b) The phonon density of states weighted by the \GP{}s, $\Gamma_i(\nu)$. 
The indices $i = 1, 2, 3, b$ correspond to $x$ uniaxial, 
$y$ uniaxial, $z$ uniaxial, and $xy$ biaxial strains, respectively.
The inset shows the densities of \GP{}s, $g_i(\nu)$, 
for $i= 1, 2, b$.
}
\label{fig:Gamma}
\end{figure}

We perform uniaxial deformations in the $x$, $y$, and $z$ directions
with strains set to $\epsilon = \pm 0.5~\%$.
For the $xy$ biaxial deformations, $\epsilon = \pm 0.25~\%$.
The \GP{}s $\gamma_{ i, \lambda\VEC{k} } $ along 
the high-symmetry directions due to these deformations are
shown in Fig.~\ref{fig:GP}(a)$-$(d). To quantify more clearly
the distribution of \GP{}s in the BZ, we calculate the density
of \GP{}s according to
\begin{equation}
g_{i}(\gamma) = \dfrac{\Omega}{(2\pi)^{3}} \sum_{\lambda}
\int_{\rm BZ}d\VEC{k}\ \delta(\gamma - \gamma_{i,\lambda\VEC{k} })
\end{equation}
which is shown in the right panels in Fig.~\ref{fig:GP}.
The density of \GP{}s due to $x$ uniaxial, $y$ uniaxial, and 
$xy$ biaxial strains are also displayed in the 
inset of Fig.~\ref{fig:Gamma}(b) for a direct comparison.
From Figs.~\ref{fig:GP}(a) and \ref{fig:GP}(d), some negative \GP{}s are observed 
near the $\Gamma$ point,
which correspond to the lowest transverse acoustic (ZA) modes.
However, the plots for densities of \GP{}s , $g_i(\gamma)$, show
most \GP{}s are populated
between the small range of $0$ to $2$, and negative \GP{}s are completely suppressed.
The $g_{i}(\gamma)$ plots also
show that large \GP{}s, say, $g_i (\gamma) > 2$,
are totally negligible when sampling
is taken. From Fig.~\ref{fig:GP}(c) we note that 
most \GP{}s are very small for $z$ uniaxial
deformation, which is consistent with the fact
that weak van der Waals interactions exist between \mos{} trilayers.

The phonon density of states of the unstrained structure is shown 
in Fig.~\ref{fig:Gamma}(a) where there is a frequency gap from
$240$ to $285$~\invcm.
The density of states weighted by the \GP{}s, $\Gamma_i(\nu)$, is shown in
Fig. \ref{fig:Gamma}(b), where a gap is inherited from Fig.~\ref{fig:Gamma}(a).
We note that, $\Gamma_3(\nu)$, which is due to a $z$ uniaxial strain, has 
a broad peak near $50$~\invcm\ due to the fact that these frequencies are 
associated with more significant \GP{}s [see Eq.~\ref{eq:Gnu}].
It is interesting to see that,
while
the density of \GP{}s due to the $xy$ biaxial strain, $g_b(\gamma)$, 
is quite different from that due to the $x$ or $y$ uniaxial
strains as shown in the inset in Fig. \ref{fig:Gamma}(b) (or even
the average of \GP{}s due to $x$ and $y$ uniaxial strains), 
$\Gamma_1(\gamma)$, $\Gamma_2(\gamma)$, and $ \Gamma_b(\gamma)$ are 
essentially the same numerically,
as shown in Fig. \ref{fig:Gamma}(b).
This justifies the proposal to use an $xy$ biaxial strain to replace
an $x$ uniaxial or $y$ uniaxial strain to 
calculate the integrated quantity $I_b(T)$.

\begin{figure}
\includegraphics[clip,width=8.0cm]{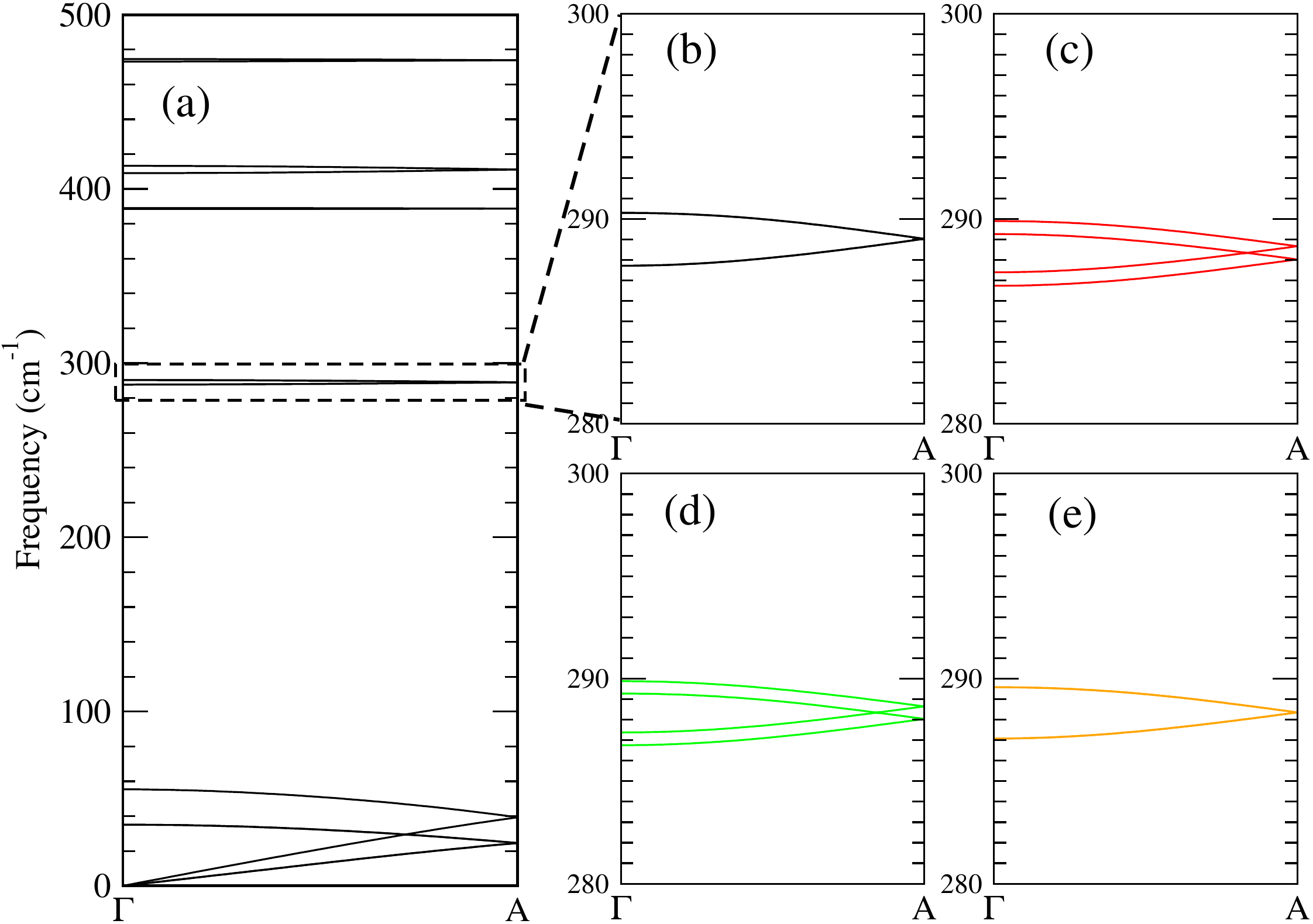}
\caption{
(Color online)
(a) The phonon dispersion of unstrained \mos{} along the $\Gamma - A$ path. 
(b) Two doubly degenerate phonon modes
for the unstrained structure (at $\Gamma$, the frequencies are
$287.7$ and $290.3$~\invcm{} for $E_{2u}$ and $E_{1g}$, respectively).  
(c), (d), and (e) 
The detailed variations of the 
phonon frequencies for the $x$ uniaxial, $y$ uniaxial, 
and $xy$ biaxial strained structures, respectively. The strain values
for uniaxial and biaxial strains are $+0.5~\%$ and $+0.25\%$, 
respectively.
}
\label{fig:ph-GA-BRGO-zoom.pdf.pdf}
\end{figure}

We now provide two pieces of evidence that explain
the difference between \GP{}s
obtained with an $xy$ biaxial strain and 
the average of \GP{}s obtained 
with $x$ and $y$ uniaxial strains.
Fig.~\ref{fig:ph-GA-BRGO-zoom.pdf.pdf} shows the phonon dispersions of the 
equilibrium structure and the strained systems along that $\Gamma-A$ path.
Figs.~\ref{fig:ph-GA-BRGO-zoom.pdf.pdf}(b)-(e) 
focus on the change in
frequencies around $290$~\invcm.
It can be seen from Figs.~\ref{fig:ph-GA-BRGO-zoom.pdf.pdf}(b) and \ref{fig:ph-GA-BRGO-zoom.pdf.pdf}(e) that
a $xy$ biaxial strain cannot destroy the degeneracies of two doubly
degenerate $E_{2u}$ and $E_{1g}$ phonon modes. However,
as seen in Fig.~\ref{fig:ph-GA-BRGO-zoom.pdf.pdf}(c),
under an $x$ uniaxial deformation, the two doubly degenerate $E_{2u}$ and $E_{1g}$ 
phonon modes around $290$~\invcm\ split into four
nondegenerate phonon modes. Fig.~\ref{fig:ph-GA-BRGO-zoom.pdf.pdf}(d)
shows the same splittings 
for a $y$ uniaxial strain.
This explains the 
average of \GP{}s due to $x$ and $y$ 
strains is not the same as that due to an
$xy$ biaxial strain. 
However, the integrations according to Eq.~\ref{eq:Gnu}
associated with 
$x$ uniaxial, $y$ uniaxial, and $xy$ biaxial
strains give rise to the same phonon density of states weighted by the Gr\"uneisen
parameters,  a fact which is expected on physical
ground.

\begin{figure}
\includegraphics[clip,width=8.0cm]{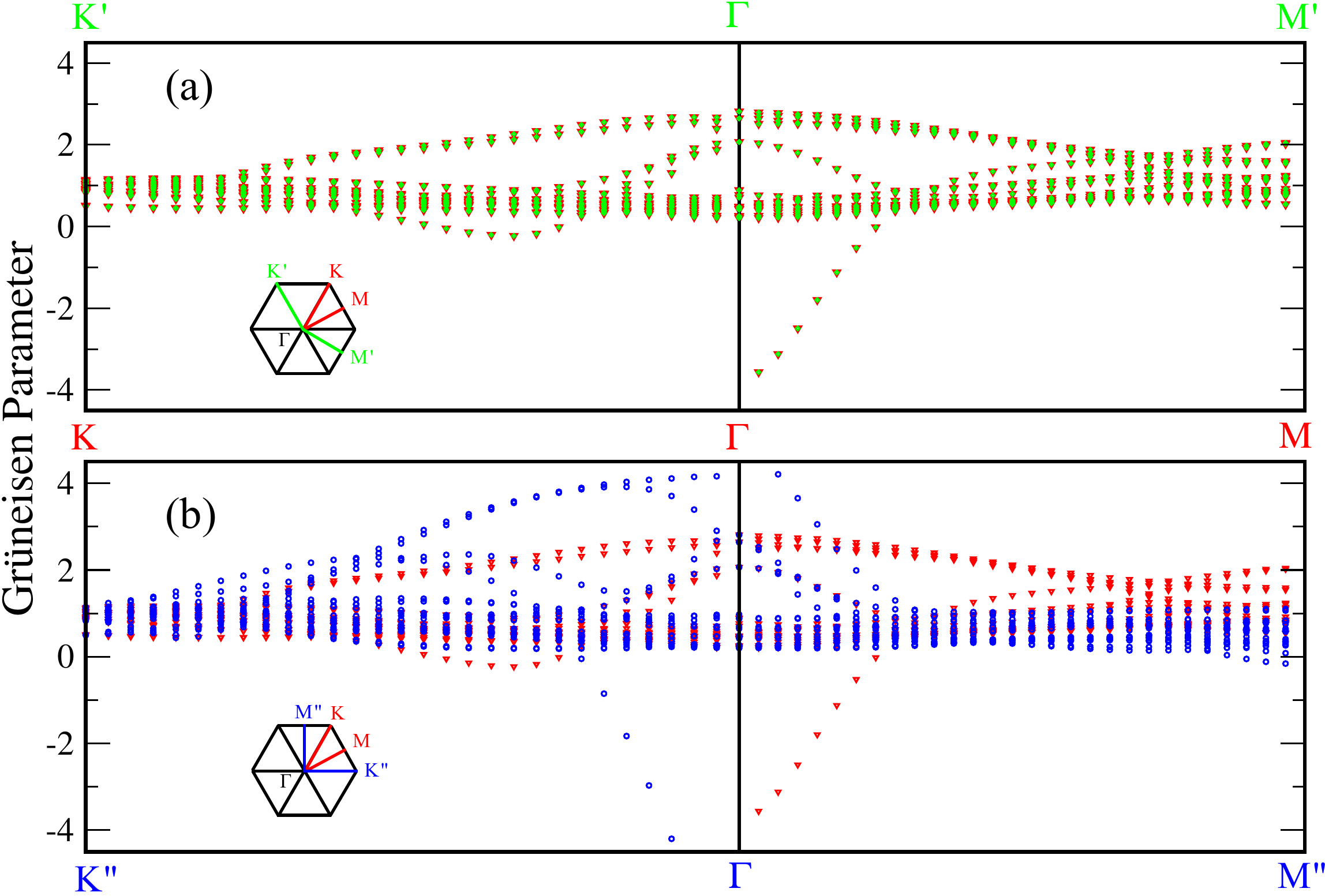}
\caption{
(Color online)
\GP{}s obtained with $x$ uniaxial strains. 
The chosen paths are (a) $K-\Gamma-M$ and $K'-\Gamma-M'$, and 
(b) $K-\Gamma-M$ and $K''-\Gamma-M''$.
The paths are shown in the insets.
}
\label{fig:KGM}
\end{figure}
Fig.~\ref{fig:KGM} furnishes
another piece of evidence that under an $x$ uniaxial strain,
the planar BZ now has
a fourfold rotation symmetry, in contrast to the sixfold rotation
symmetry for the case of biaxial strain (where the hexagonal symmetry
is preserved).
The agreement of \GP{}s 
along $K-\Gamma$ and $K'-\Gamma$ paths in
Fig.~\ref{fig:KGM}(a) shows that there
is a reflection symmetry around the $y$ axis. Similarly, the agreement 
of \GP{}s between those along the $\Gamma-M$ and $\Gamma-M'$ paths shows that there
is a reflection symmetry around the $x$ axis. 
Under the sixfold rotation symmetry of the planar BZ, we expect \GP{}s to be the same 
along the $K-\Gamma$ and $K''-\Gamma$ paths, 
or along the $\Gamma-M$ and $\Gamma-M''$ paths. However, Fig.~\ref{fig:KGM}(b)
shows that there is no agreement of \GP{}s 
along the $K-\Gamma$ and 
$K''-\Gamma$ paths,  or along the $\Gamma-M$ and $\Gamma-M''$ paths.

\begin{figure}
\includegraphics[clip,width=8.0cm]{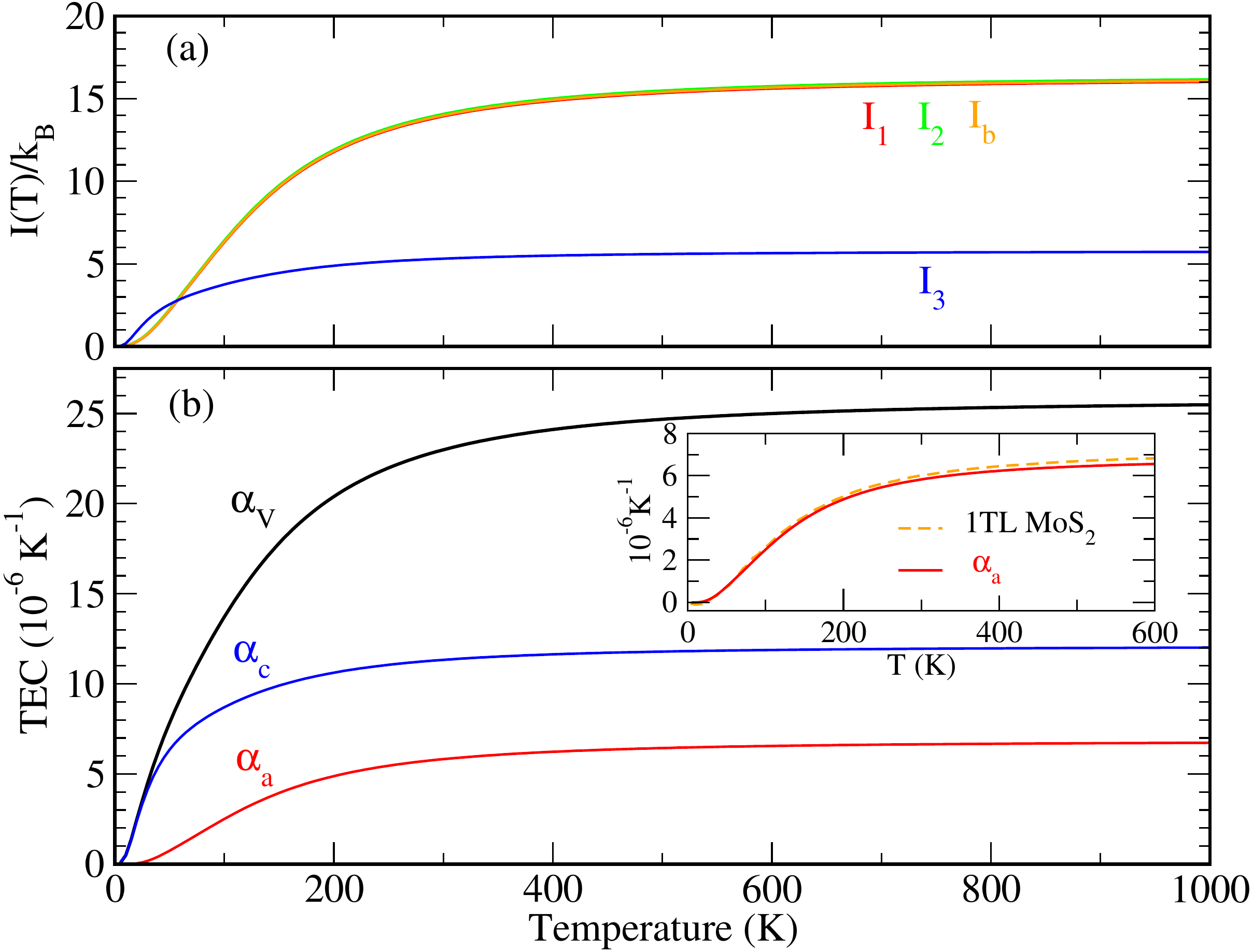}
\caption{
(Color online)
(a) The integrated quantities $I_i(T)$ as a function of temperature.
(b) The linear TECs of \mos{} along the $a$ and $c$ directions denoted by $\alpha_a$ and $\alpha_c$, respectively, as
a function of temperature.
The volumetric TEC is denoted by $\alpha_v$. The insert shows the comparison of $\alpha_a$ with 
the in-plane TEC of a single-trilayer \mos{}, obtained using a QHA approach\cite{Sevik14v89}.
}
\label{fig:ltec}
\end{figure}

The integrated quantities $I_1(T)$, $I_2(T)$, and $I_b(T)$ shown in Fig. \ref{fig:ltec}(a) 
are essentially identical, and they are much larger than $I_3$, which is due to 
a $z$ uniaxial strain. This may be traced to the fact that interactions between \mos{} trilayers are weak compared
to in-plane interactions that result in smaller \GP{}s (i.e.,
smaller frequency changes) for $z$ strains.
The linear TECs in the $a$ and  $c$ directions are shown in Fig. \ref{fig:ltec}(b).
Even though $I_{3}(T)$ is about three times smaller 
than $I_{b}(T)$ for, say, $T>400$~K, 
$\alpha_{c}(T)$ is larger than $\alpha_{a}(T)$. At $T = 1000$~K, $\alpha_a = 6.73 \times 10^{-6} $~K$^{-1}$, 
$\alpha_c = 12.01 \times 10^{-6} $~K$^{-1}$, and the volumetric TEC $\alpha_v = 2\alpha_a + \alpha_c = 25.47 \times 10^{-6}$~K$^{-1}$.
The main reason for this is that the value of $C_{33}$ ($51.70$~GPa) 
is much smaller
than the value of $C_{11} + C_{12}$ ($301.19$~GPa); therefore,
according to Eq.~\ref{eq:2d1}, it is possible that $\alpha_c$ is 
larger than $\alpha_a$. 
The result that $\alpha_c$ is indeed larger than $\alpha_a$ is consistent
with the physical fact that it is easier to perform a deformation in the $z$ direction
than in the in-plane direction, which is again, attributed to the weak 
interactions in the out-of-plane direction.
This is also confirmed by
the results shown in the inset of Fig. \ref{fig:ltec}(b) where there is a striking similarity between the temperature dependences
of in-plane TECs for both the bulk \mos{} and a single-trilayer \mos{}, which is obtained from a QHA-LDA treatment\cite{Sevik14v89}. 
El-Mahalawy and Evans\cite{El-Mahalawy76v9} measured the linear TECs of 2H-\mos{} between $293$ and $1073$~K and 
reported $\alpha_{a} =
1.9 \times 10^{-6}$~K$^{-1}$ and $\alpha_{c} = 8.65 \times 10^{-6}$~K$^{-1}$, which are consistently lower than our values.
The same group\cite{Murray79v12} again measured the TECs of 2H-\mos{} between $10$ and $320$~K and
found a larger $\alpha_{a} = 4.922 \times 10^{-6}$~K$^{-1}$, which agrees better with our result, and
$\alpha_{c} = 18.580 \times 10^{-6}$~K$^{-1}$, which
is somewhat larger. 
We find the behavior of the rate of change of $\alpha_c$ with $T$ to be
different from that of $\alpha_a$, where the former increases rapidly at low $T$
while the latter increases more gradually with $T$.
Murray and Evans\cite{Murray79v12}
have pointed out that the lattice constant $a$ ($c$) increases linearly (nonlinearly)
with temperature, which is consistent with our TEC results at low $T$.

\section{Conclusions}
\label{sec:conclusions}
In summary, we have proposed a direct way to calculate the linear thermal expansion
coefficients (TECs) of a hexagonal system based on the Gr\"uneisen formalism. 
We have also proposed a way to replace the inefficient symmetry-lowering uniaxial strains by 
the efficient symmetry-preserving biaxial strains.
We successfully implemented the computational schemes and applied them to a technologically important
material, \mos{}. 
We found that \mos{} has a large TEC anisotropy where 
the thermal expansion coefficient in the $c$ direction is $1.8$ times 
larger than that in the $a$ direction at high temperatures.
We highlighted that even though the integrated quantities $I_i(T)$ required 
by the TEC calculations can be obtained via a symmetry-preserving biaxial strain, the
\GP{}s from a biaxial strain
may not be a simple average of the \GP{}s from uniaxial $x$ and $y$ strains. 
We demonstrated that we only need a minimum of two symmetry-preserving deformations
to directly calculate the 
TECs of a general hexagonal system.
In contrast, the quasiharmonic approximation, when dealing with a two-parameter system, may require an
expensive search in the two-dimensional search space. Therefore, we expect that the strategies adopted in this paper
to treat a general two-lattice-parameter hexagonal system
can be similarly applied to treat other two lattice-parameter systems such as trigonal and tetragonal systems, 
thus opening the door for a truly predictive TEC calculation for many important materials. 
We also expect the TEC calculations based on the Gr\"uneisen
formalism via symmetry-preserving deformations may be readily incorporated in any phonon related codes
such as \textsc{phonopy}\cite{Togo15v108}.

\section*{Acknowledgments}
We acknowledge stimulating and fruitful discussions with Ching Hua Lee.
We thank the National Supercomputing Center, Singapore,
for computing resources.
Y.Y.F.L. acknowledges 
support from the Singapore National Science Scholarship.
\bibliographystyle{plain}

\bibliographystyle{apsrev}

\end{document}